# Low Temperature Behaviour and Glass Line in the Symmetrical Colloidal Electrolyte

Jose B. Caballero[1, *] and Antonio M. Puertas[1]

[1] *Group of Complex Fluids Physics, Department of Applied Physics, University of Almeria, 04120 Almeria, Spain*

We report on the low temperature behaviour of the colloidal electrolyte by means of Molecular Dynamics simulations, where the electrostatic interactions were modeled using effective screened interactions. As in previous works, we have found a region of gas-liquid coexistence located in the low $T$- low $\rho$ region. At temperatures much lower than the critical one, the system cannot reach equilibrium, that is, the gas-liquid transition is arrested. Two different mechanisms have been identified to cause arrest: gelation in the lowest $T$ and crowding at intermediate $T$ values, the latter associated with the crossing point between the binodal and the glass line. To test the latter, the dynamics of the colloidal electrolyte near this crossing point has been computed and compared to the universal predictions of the ideal MCT. As in other glass-forming liquids, we found good agreement between this mean field theory and the dynamics of this complex system, although it fails just at the transition. Interestingly, in this region we found that the dynamics of this system is driven mainly by the steric interactions, showing all the typical properties of a repulsive colloidal glass. Finally, the isodiffusivity lines show that in this system with short range attractions, there is no re-entrant glass phenomenon, as opposed to monocomponent systems.

## I. INTRODUCTION

Colloidal electrolytes have been introduced recently as the colloidal analog of the Restricted Primitive Model (RPM) for ionic systems; i.e. a 1:1 mixture of equally sized, equally charged colloids bearing opposite surface charges. The main difference between colloidal and ionic electrolytes is not the size, but the ability to control easily the range of the interactions in the former, the system being more complicated and interesting than just a 'big ionic electrolyte' [1]. Different crystal phases have been observed at high density [2–4], which makes them particularly interesting for potential applications in photonics and other material sciences [5]. On the other hand, experiments have revealed that some properties of proteins can be also reproduced in colloidal electrolytes [6, 7].

The equilibrium phase diagram of colloidal electrolytes is under study both experimentally [2–4, 6, 7] and by computer simulations [8–12]. In addition to the variety of crystal phases found, most of them reproduced by the simulations and theoretical calculations, a liquid-gas transition has been predicted by simulations in the low temperature – low density – long interaction range region where strong correlations between oppositely charged particles are dominant. The critical temperature of this transition presents a non-monotonic behaviour as a function of the interaction range, that is, the salt concentration [9], contrary to monocomponent systems [13] (although a monotonic trend can be recovered assuming the DLVO effect of the salt concentration on the contact value of the interaction potential [12]). The liquid phase is restricted to a narrow range of temperatures (between the critical and the triple one), but its stability as a function of the interaction range is still matter of debate [10, 12]. In addition to all these phenomena, recent experiments have found arrested liquid-gas separation and gels at low salt concentrations, i.e. strong, long range interactions [7] (gels were defined, in analogy with monocomponet systems, as amorphous, low density solids [14]).

As shown by Sastry [15], a liquid-gas transition can be impeded by a glass transition crossing the liquid branch of the coexistence line. The liquid would then glassify and the phase separation would be arrested. On the other hand, it has been shown recently that in colloidal systems with short range attractions a new glass transition appears at low temperature induced by long lived bonds [16–21]. This attraction-driven glass transition connects continuously with the repulsion-driven glass transition, usual in atomic and molecular systems and in hard spheres, at high density. Depending on the range of the interaction, upon decreasing the temperature a glass-fluid-glass re-entrance phenomenon can be produced [16, 18, 22] and a high order singularity [18, 19, 23] can appear in the phase diagram, interacting with the phase transitions. This scenario with two glass transitions was anticipated by Mode Coupling Theory [18, 24–26], and later confirmed by experiments and simulations [16, 27]. The relation between the attraction driven glass transition and gelation, however, is as yet unclear due to the interplay of the phase transition and the two glasses [28–30].

In this paper, we address the low temperature behaviour of several colloidal electrolytes with different interaction ranges by means of Molecular Dynamics simulations [31]. As in previous works, the electrostatic interactions were modeled using a screened interaction model. Simulations along two different isochores were carried out from super critical temperatures down to temperatures deep in the coexistence region. A polydisperse system is used in order to avoid crystallization, and expand liquid phase to higher density and lower temperatures (i.e. to the glass transition). Just below the critical point, we find gas-liquid separation as expected, but the separation is arrested at lower temperatures. We have observed two mechanisms causing the arrest: crowding of the liquid, at moderate temperatures, and gelation for very low ones. The former is associated with the crossing between the binodal line and the glass line and is studied in detail in this work. Precisely, the dynamics in this region has been computed and compared with the universal predictions of the Mode-Coupling Theory [32]. We find a rather good agreement with the theory, as in other glass-forming liquids [33, 34]. The properties of the transition indicate that it is driven by the core-core repulsion, and not by the attractions which induce liquid-gas separation, for all ranges studied. Moreover, we determine the iso-diffusivity lines which serve as an estimation of the glass line, which shows no signature of glass-liquid-glass re-entrace for any interaction range, in agreement with the driving mechanism of the transition and in spite of the attractions in the system.

*Corresponding author:jcaballe@ual.es

The paper is organized as follows: in Section II, we present the model and the simulations details are explained – in particular the interaction potential is discussed. Section III is devoted to results and discussions: first the arrest of the phase separation is shown, then we study the dynamics in the vicinity of the crossing point between the glass line and the liquid branch, and finally the iso-diffusivity lines are shown. We conclude in Section IV presenting the main conclusions.

## II. MODEL AND SIMULATIONS

The scope of this work, as it has been already mentioned, is to study the low temperature behaviour (at moderate densities) and the collective dynamics of the symmetrical colloidal electrolyte, that is, a binary mixture of $N$ spherical colloidal particles of equal diameter (immersed in a continuous medium characterized by its dielectric constant $\varepsilon$); $N/2$ bearing a surface potential $+\phi$ and $N/2$ with $-\phi$. Previous computer simulations focused on the equilibrium phase behaviour modeling the electrostatic interaction by means of a linear screening theory (concretely the DLVO model, developed by Derjaguin and Landau, and Verwey and Overbeek [35]) and short-ranged steric repulsions with hard-core interactions. This system can thus be considered as the colloidal analogue of the widely studied Restricted Primitive Model (RPM). Obviously this is an approximation to the whole problem, where the ions are not simulated. However, the global problem, where the small ions are considered, requires a big amount of computational time and the coexistence lines cannot be easily obtained [36].

Since we are concerned with the dynamical aspects of this system, we will carry out Molecular Dynamics (MD) simulations where the Newton equations are numerically solved using a Verlet algorithm [31]. There are two important aspect to be taken into account. First, Brownian Dynamics (BD) simulations should be performed to reproduce the microscopic dynamics of the real system. Nevertheless, it has been proved that the collective dynamics of the whole system does not depend on the microscopic details [37], and since the relaxation time is shorter for Newtonian systems, MD algorithms were employed throughout. The second point is related with the potentials, since MD cannot be applied directly to discontinuous potentials (event driven algorithms have been developed for hard spheres or square wells). Therefore the potential used in previous works has been slightly modified by changing the steric interactions for a soft potential, $u(r) \sim 1/r^{36}$. Now the total interaction is given by: $u_{total} = (\sigma_{12}/r)^{36} \pm \psi exp[-\kappa(r-\sigma_{12})]$. However, this manipulation introduces two non desirable effects in the minimum of the attractive interactions (see inset in Fig. 1): *i)* it is displaced toward longer distances and *ii)* its value changes considerably. Thereby, a fourth-order polynomials has been included in such a way these effects are minimized (see Fig. 1). The whole potential thus reads as follows:

$$U(r) = \begin{cases} (\sigma_{12}/r)^{36} + A(r-\sigma_{12})^3 + B(r-\sigma_{12})^2 & \\ \quad +C(r-\sigma_{12}) + D & r \leq R_a \\ \pi\varepsilon\sigma_{12}\phi_1\phi_2 \cdot \exp\{-\kappa(r-\sigma_{12})\}, & r > R_a \end{cases} \tag{1}$$

where $\sigma_{12}$ is center to center distance between particles 1 and 2, $\phi_1$ and $\phi_2$ are the surface potentials, and $\kappa$ is the inverse Debye length, which depends on the ionic concentration. Four conditions were imposed to determine the values of the parameters $A$, $B$, $C$, and $D$: $U(r=\sigma)=0$, $U_{min}=-1$ and continuity for the potential and its first derivative (see lower inset in Fig. 1). The values for $R_a$ were set to reproduce the results presented in ref. [9]: $R_a = 1.05$ for $\kappa\sigma = 6$ and $\kappa\sigma = 10$; and $R_a = 1.02$ for $\kappa\sigma = 20$. To avoid crystallization below the triple point [10], we used polydisperse samples, where the diameters were taken from a flat distribution with mean diameter $\sigma$ and width $w = 0.1\sigma$ for $\kappa\sigma = 6$ and $\kappa\sigma = 10$, and $w = 0.15\sigma$ for $\kappa\sigma = 20$. Hereafter, reduced units will be used: $\sigma = 1$ (the mean diameter), $U^* = U/(\pi\varepsilon\sigma\phi^2)$ and so $T^* = k_BT/(\pi\varepsilon\sigma\phi^2)$ defines the reduced temperature, the density is defined as $\rho^* = N\sigma^3/V$ (where $N$ is the number of particles, and $V$ is the volume of the system), and finally, the time is measured as $t^* = \sigma\sqrt{m/(\pi\varepsilon\sigma\phi^2)}$, $m = 1$ being the mass of the particles.

Constant temperature MD simulations were performed along two isochores $\rho^* = 0.25$ and $\rho^* = 0.05$ (for $\kappa\sigma = 6$, $\kappa\sigma = 10$, and $\kappa\sigma = 20$) from $T = 0.19$ (super critical state) to $T^* = 0.01$ (well deep in the gas-liquid coexistence region). For each temperature, the system is equilibrated at $T^* = 5$ and then is instantaneously quenched at $t^* = 0$ to the desired temperature. Three different quenches were performed at each temperature for statistics: the first two with Andersen thermostat [31] and another one rescaling the particle velocities at regular time steps. The static properties of the model were found independent of the thermostat used. For each simulation, we monitored the time evolution of the energy per particle, a global orientational order parameter, $Q_6$ (which presents values close to zero in the liquids and non zero values in the crystal phases [38]), and a parameter to account for the density homogeneity, defined as:

$$\Psi_4 = \frac{1}{64\rho^{*2}} \sum_{j=1}^{64} (\rho^* - \rho_j^*)^2 \qquad (2)$$

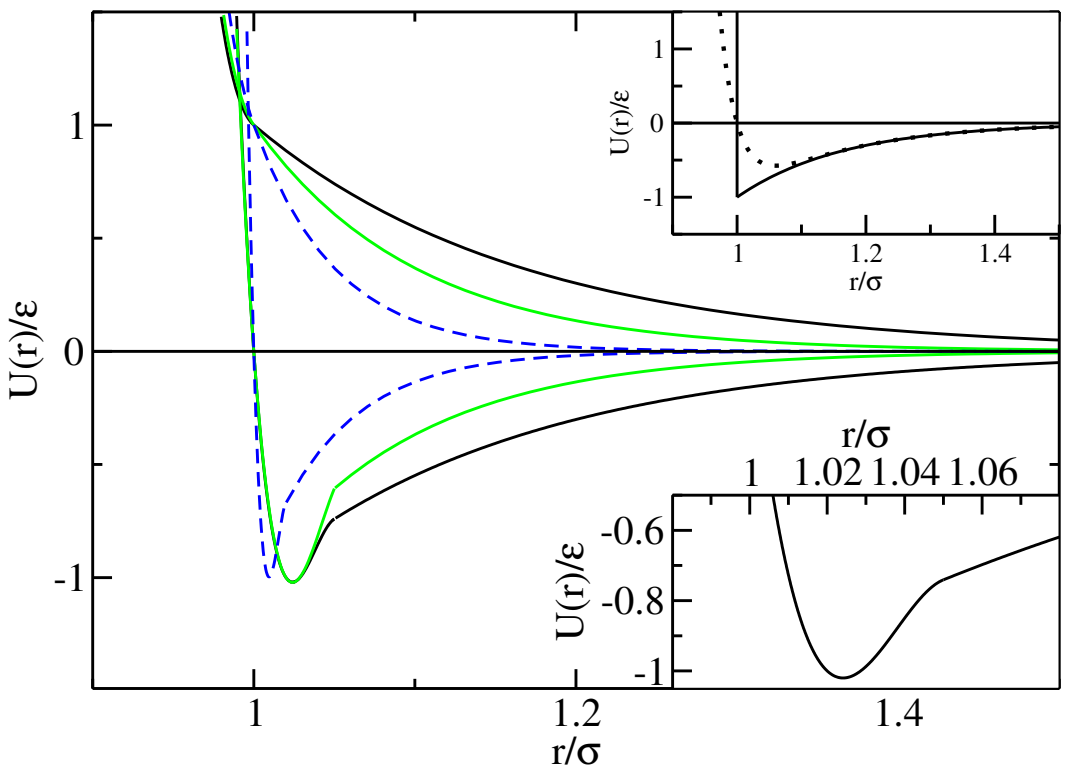

FIG. 1: Interaction potentials for different $\kappa\sigma$ values: $\kappa\sigma = 6$ (black lines), $\kappa\sigma = 10$ (grey lines) , and $\kappa\sigma = 20$ (dashed lines); where repulsive and attractive interactions are shown. Upper inset: comparison between the colloidal electrolyte with hard-spheres (solid line) and soft-spheres (dotted line) for $\kappa\sigma = 6$. Lower inset: enlargement near $R_a$ of the attractive potential with $\kappa\sigma = 6$.

Here the simulation box is divided into 64 equal portions with densities $\rho_j^*$, while $\rho^*$ denotes the density of the whole system. $\Psi_4$ will be close to zero in homogeneous phases, while it presents higher values in non-homogeneous states. Simulations comprise, at least, 20000 reduced time units, where the time step was set to $\delta t^* = 0.001$.

## III. RESULTS

### A. Low Temperature Behaviour

Figure 2 shows instantaneous configurations at different temperatures along the aforementioned isochores (i.e. $\rho^* = 0.25$ and $\rho^* = 0.05$) for $\kappa\sigma = 6$. It is worth observing the temperature evolution. At the highest temperature, the system remains homogeneous with large density and energy fluctuations what means it is close to its critical point. As $T^*$ decreases, the system undergoes a gas-liquid transition ($T^* = 0.15$ and $T^* = 0.09$), consistent with the results obtained by Monte Carlo simulations in ref. [9]. Finally, at lower $T^*$ values the system becomes ramified, $T^* = 0.05$ and $T^* = 0.01$, and the transition does not occur. Just at the lowest temperature investigated ($T^* = 0.01$) a percolating cluster containing all the particles is formed, where the particles are restricted to move only sampling a very small portion of the box, that is, they are completely caged. In analogy with the literature in monocomponent model systems, we identify this spanning network as a *gel* [39–43]. Note that gelation, as in monocomponent systems, is density dependent because it does not appear for $\rho^* = 0.05$ at $T^* = 0.01$, though it was found again at even lower temperatures. For the symmetrical colloidal electrolyte, gelation is consistent with the experimental results at high interaction strengths [7] (there obtained at low salt concentrations). The dynamics aspects of the gel will be studied elsewhere.

In addition to direct observation, we have estimated the homogeneity parameter $\Psi_4$ during the simulations. In Figure 3 the temperature dependence for the long time limit of $\Psi_4$ is shown (circles for $\kappa\sigma = 6$). In equilibrium, one expects that $\Psi_4$ increases its value as the temperature is decreased when a gas-liquid transition is present (the lower temperature, the denser the liquid). Indeed, we find such a behaviour at high temperatures, that is, the behaviour is consistent with equilibrium expectations. However, $\Psi_4$ develops a clear maximum, at $T = 0.11$, and thereafter decreases monotonically up to values close to those present in super critical states, opposite to the expected equilibrium evolution. The temperature axes can thus be divided into two regions: *equilibrium* at high enough $T^*$ and *non equilibrium states* characterized by an anomalous evolution of $\Psi_4$. Moreover, looking into non-equilibrated systems ($T^* = 0.09$, $T^* = 0.05$, and $T^* = 0.01$ in Fig. 2), we observe that the maximum cannot be uniquely related with the formation of ramified dense phases, but also there exist compact dense phases (associated with the formation of a liquid) which cannot be equilibrated throughout the simulation. Thus, the maximum is not only associated to gelation, but also with crowding.

To gain insight in the phenomenology arising in our system, we performed simulations for the binary mixture of Lennard-Jones particles introduced by Kob et al. [45] calculating $\Psi_4$ along its critical isochore (inset in Fig. 3)– the critical parameters for this system are: $T_c^* = 1.2$ and $\rho_c^* = 0.3$. In this case, $\Psi_4$ also shows a maximum at a lower temperature than the critical one, resembling the behaviour found in our system. Interestingly, the temperature where the maximum occurs, coincides with the crossing point between the binodal and the glass line estimated by Sastry, $T^*_{Sastry} = 0.16$ [15] (in our simulations $T^*_{cross} = 0.15 \pm 0.05$). Therefore, we associate this maximum with the gas-liquid transition arrested by the glass line.

In analogy with the Lennard-Jones system, we identify the maximum in the colloidal electrolyte as the crossing point between the gas-liquid transition and the glass line.

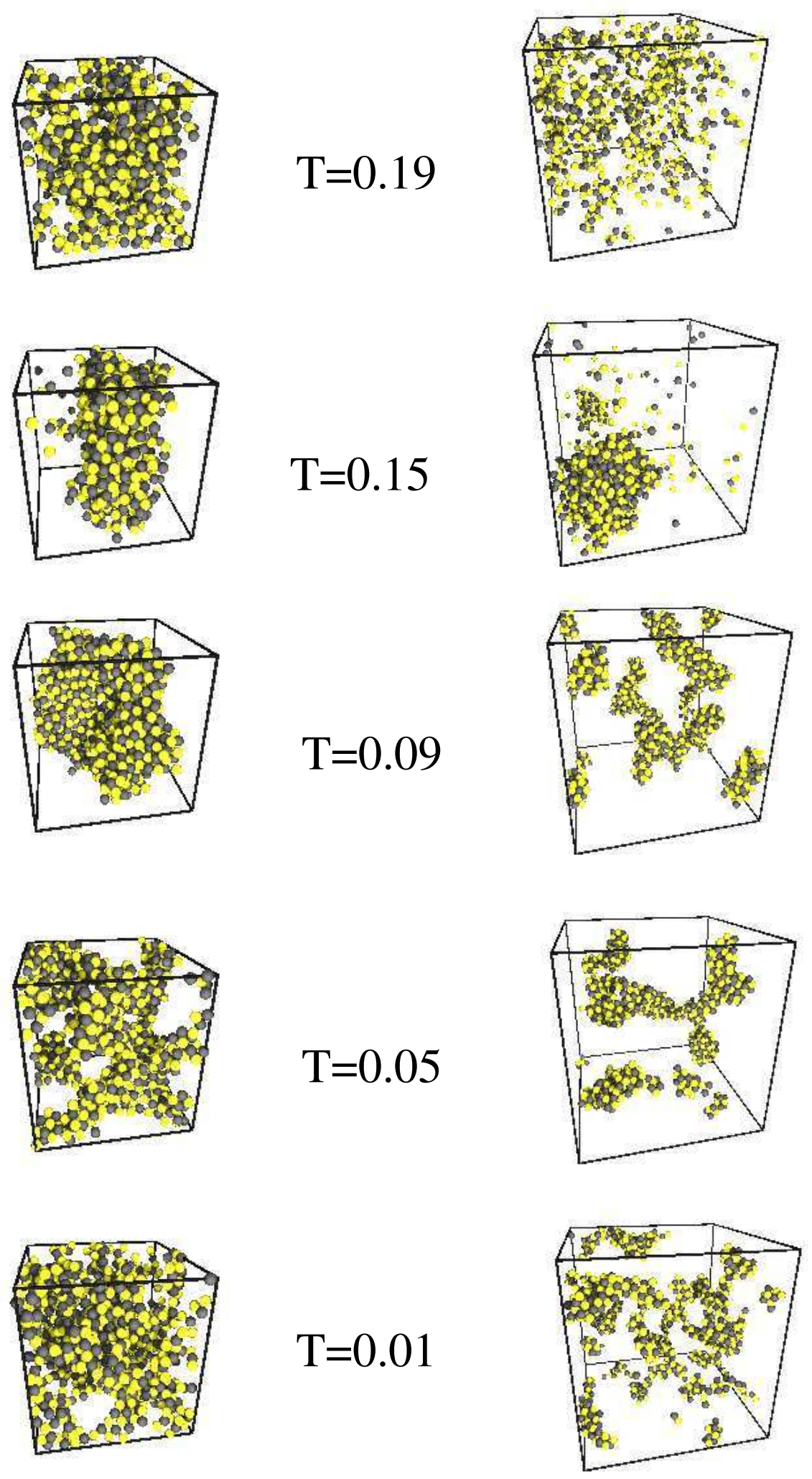

FIG. 2: Instantaneous configurations for the colloidal electrolyte with $\kappa\sigma = 6$ at two different ishocores, $\rho = 0.25$ (left) and $\rho = 0.05$ (right): $T = 0.19$ (super critical temperature), $T = 0.15$, $T = 0.09$, $T = 0.05$, and $T = 0.01$.

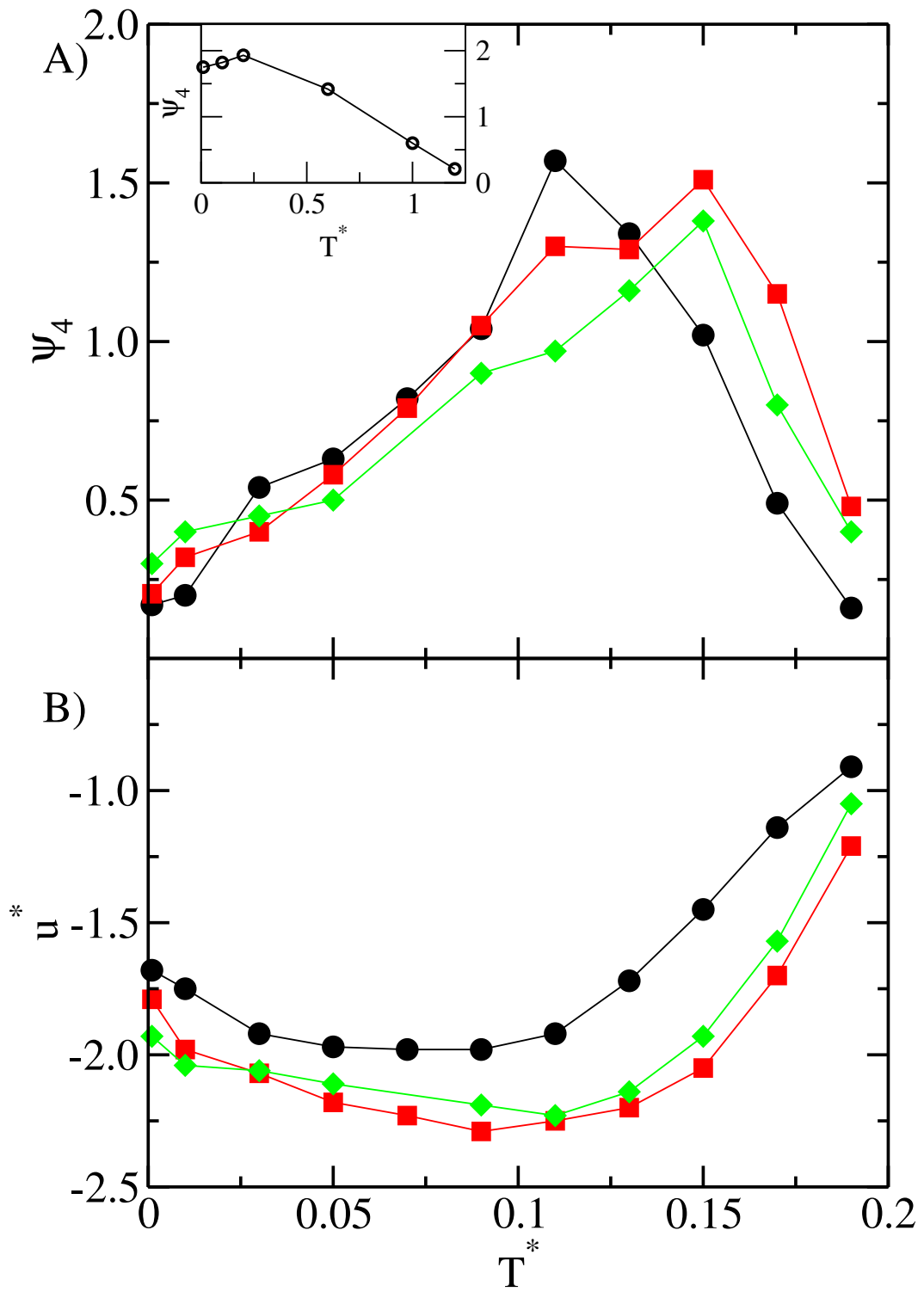

FIG. 3: A) Temperature dependence of the homogeneity parameter at $\rho = 0.25$ for: $\kappa\sigma = 6$ (circles), $\kappa\sigma = 10$ (squares), and $\kappa\sigma = 20$ (diamonds). Inset: Homogeneity parameter for a Lennard-Jones mixture along the isochore $\rho = 0.3$. B) Energy per particles versus the temperature for the same states as in the upper panel.

Furthermore, it is worth observing how $\Psi_4$ diminishes weakly for the Lennard-Jones mixture and it decreases abruptly in the case of the colloidal electrolytes, where gelation takes place. Thus, we conclude that the gas-liquid transition is arrested by two different mechanisms: *i) crowding* at high enough temperatures (where the liquid becomes a glass); and *gelation* in low temperature regimes.

Similar behaviours were obtained for $\kappa\sigma = 10$ and $\kappa\sigma = 20$ (Fig. 3), that is: super critical states at high temperatures, gas-liquid phase separation, and arrested phase separation by crowding or gel formation. Note that the aforementioned picture is not clear from the energy curves, where the temperature variation is softer. One important finding is that the colloidal electrolyte forms a gel at low temperatures even when the interaction range is moderately long (for example $\kappa\sigma = 6$). This is opposite to monocomponent model systems, where gel formation is associated to short-ranged interactions [39–43].

The presence of gels at moderate interaction ranges can be rationalized with the concepts used in electrolyte solutions. In the symmetrical ionic fluid, screening between adjacent ions is fundamental to understand its critical behaviour: although the bare Coulomb force is extremely long ranged, the *effective* interaction between any two particles of an electrolyte is screened by the other ions and therefore is short ranged. Consequently, the most recent simulation works indicate that the critical behaviour of the RPM belongs to the 3D-Ising universality class [44], which is the same as for liquids with short interactions. In the same way, gels in colloidal electrolytes at moderate interaction ranges arise since the *effective* interaction is shortened due to the presence of other colloids with opposite charge in the vicinity of a central particle.

## B. High Density Dynamics

Next, we analyze the dynamical properties of the colloidal electrolyte close to the crossing point of the liquid branch with the glass line. Whereas the temperature of this state was directly obtained from Fig. 3, in order to estimate its density we computed the local density for each particle in states showing liquid-vapour separation, near the crossing point (with higher temperatures). Then, the histogram of local densities was built and the *liquid density* was assumed as the density of the maximum for such a histogram. For $\kappa\sigma = 6$ we obtained: $T_{crossing} = 0.10 \pm 0.01$ and $\rho^*_{crossing} \approx 0.90$. Thereby, simulations at constant temperature (rescaling the velocities) were performed along two distinct isochores: $\rho^* = 0.85$ and $\rho^* = 0.90$. Since the results for all the isochores were qualitatively similar, we will focus our attention on $\rho^* = 0.85$ with $\kappa\sigma = 6$ where a wide temperature range was covered: $T^* = 1.0$, $T^* = 0.50$, $T^* = 0.30$, $T^* = 0.20$, $T^* = 0.18$, $T^* = 0.16$, $T^* = 0.14$, $T^* = 0.13$, $T^* = 0.125$, $T^* = 0.115$, $T^* = 0.11$, and $T^* = 0.10$. The system did not reach a stationary state in energy or dynamics for the last temperature in our time scale ($T^* = 0.10$), but the rest of the states are well equilibrated, homogeneous and fluid-like throughout the time when the measurements were taken. As expected, the dynamics is equal for positive and negative colloids, so the results will be presented using global quantities hereafter.

In Figure 4 we show the time dependence of the mean squared displacement along this isochore. At short times, $< \Delta r^2(t) > \sim t^2$ since we are in the ballistic regime; while at long times, the behaviour is diffusive. This change is due to the collisions of the particle with its neighbours that surround it at time $t = 0$. For low temperatures, both regimes are well separated by a plateau which is present at intermediate times. This plateau is due to the cage effect, i.e. the temporary trapping of the particle by its neighbours. This two-step relaxation scenario is also visible in the self part of the density correlation function $F^s(q,t) = \frac{1}{N} < \sum_i exp[-i\vec{q}(r_i(\vec{0}) - r_i(\vec{t}))] >$ (also called incoherent intermediate scattering function). A first relaxation process, the $\beta$ relaxation, occurs at short times, due to particles exploring the cages formed by their nearest neighbours. A second relaxation, the $\alpha$ relaxation, takes place at longer time scales, when particles are able to escape their cages. This complex relaxation as well as the abrupt slow down of the dynamics, are typical indications of a system approaching a glass transition, in analogy to another glass-forming liquids [34].

Since Mode-Coupling Theory (MCT) [32] is able to predict correctly many of the dynamical properties of different systems upon approaching a glass transition [33], we have compared the dynamical properties of the colloidal electrolyte with the universal prediction of this theory. In this case, the glass line is understood as a dynamical arrest of the density fluctuations due to the coupling of density modes. First of all, we point out that the comparison will be carried out with the universal predictions known for monocomponent systems, because no two component MCT calculations have been performed for the colloidal electrolytes. We begin testing the time-temperature superposition principle which states that at

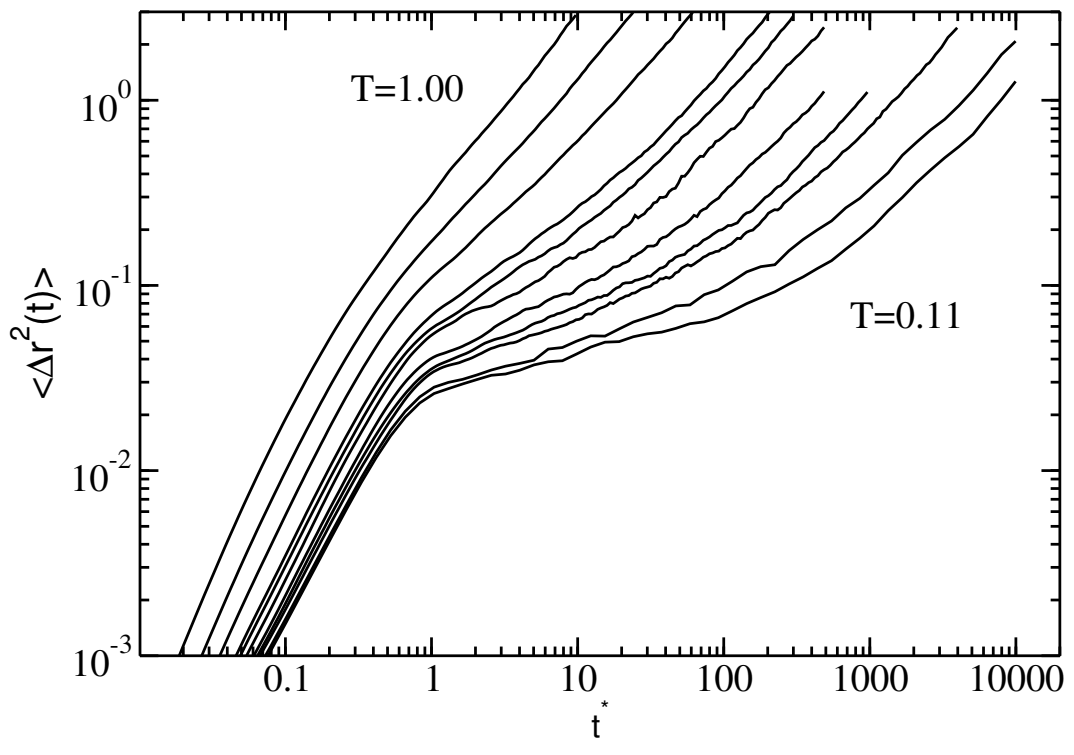

FIG. 4: Time dependence of the mean squared displacement at $\rho = 0.85$ ($\kappa\sigma = 6$): $T = 1.0$, $T = 0.50$, $T = 0.30$, $T = 0.20$, $T = 0.18$, $T = 0.16$, $T = 0.14$, $T = 0.13$, $T = 0.125$, $T = 0.115$, and $T = 0.11$ (from left to right).

a temperature $T$ the time dependence of a correlator for the $\alpha$-decay can be written as follows:

$$F^s(q,t) = \phi^s_q(t/\tau^s(T)) \quad (3)$$

where $\phi^s(x)$ is a master curve and $\tau^s(T)$ is the $\alpha$-relaxation time at $T$. The meaning of this equation is that the $\alpha$-relaxation can be described by a function which is independent of the temperature, once the time has been scaled properly. In Fig. 5 the incoherent intermediate scattering functions are plotted as a function of $t/\tau^s$, where $F^s(q,\tau^s) = 1/e$, along the aforementioned isochore. Note that below $T = 0.20$ the correlators fall onto a single master curve in the $\alpha$-decay regime, as the MCT predicts. It is worth noting that this prediction is fulfilled without any fitting parameter. The long time behaviour of the correlators can be described by a simple exponential at high temperatures. On the contrary, at low temperatures the master curve is well described by a Kohlsrausch function (dashed line in this plot), often called stretched exponential:

$$F^s(q,t) = A^s_q exp[-(t/\tau^K_q)^{\beta_q}] \quad (4)$$

here $A^s_q$ is an amplitude, $\tau^K_q$ is the relaxation time, and $\beta_q$ is the stretching exponent. As in other glass-forming liquids, we have obtained an exponent $\beta_q < 1$. The reason why the time correlation function shows at low $T$ a non-Debye behaviour is still matter of debate [46]. This stretched exponential also fits the numerical solutions of the MCT equations for monocomponent systems [32].

According to the ideal MCT, the non-ergodic transition is due to the arrest of the density fluctuations, producing the divergence of the $\alpha$-relaxation at finite temperature:

$$D^{-1} \propto \tau^s \sim (T - T_{MCT})^{-\gamma} \quad (5)$$

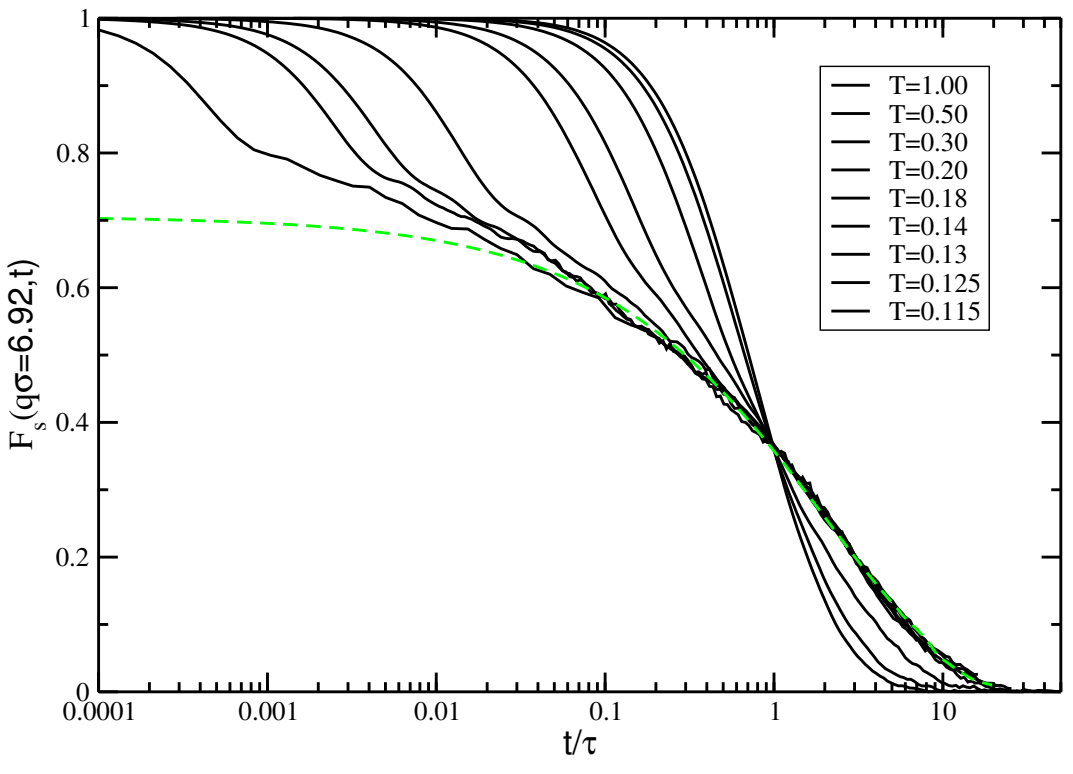

FIG. 5: Incoherent density correlation function versus $t/\tau^s(T)$, for the same states as in Fig. 4, at $q\sigma = 6.92$ (where $S(q)$ shows its maximum value). $\tau(T)^s$ is the $\alpha$-relaxation time of the system. Dashed line is the *Kohlrausch* fit.

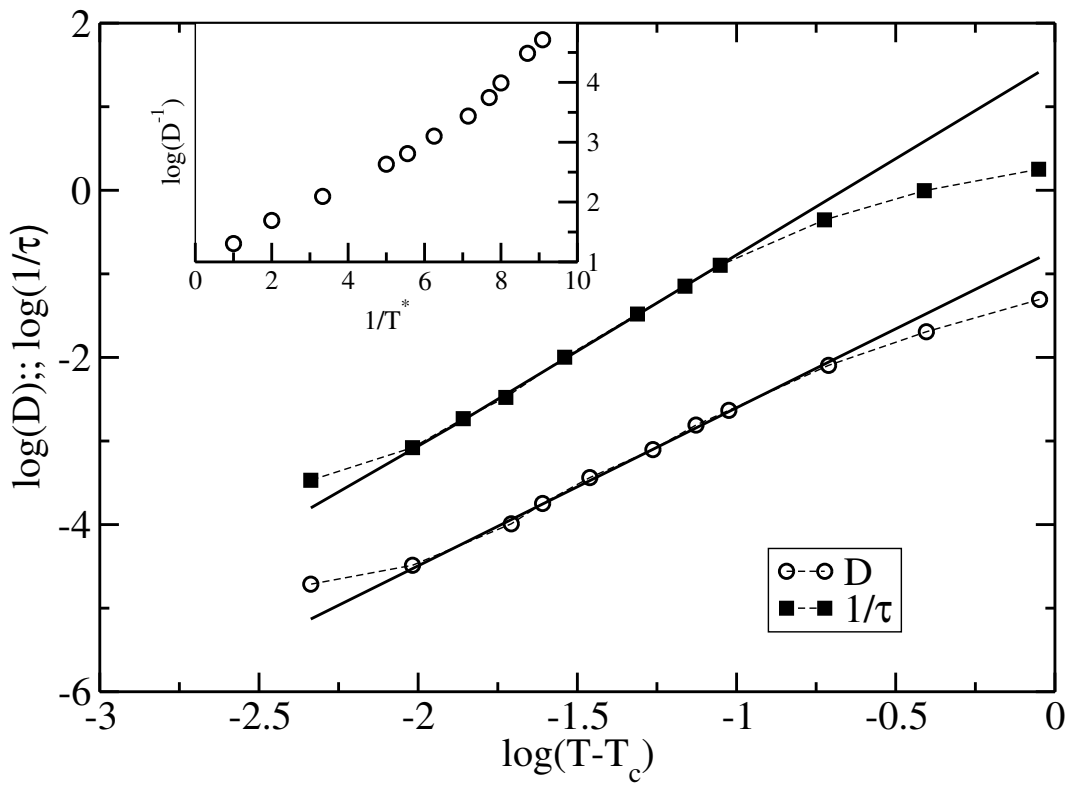

FIG. 6: Diffusion constant (filled squares) and inverse of the $\alpha$-relaxation time (open squares) versus $T - T_c$ (log scales are used) for the same states as in Fig. 4. The bold straight lines are fits to the data with eq. 5. Inset: Arrhenius representation of the diffusion constant along this isochore.

We can verify this prediction from our simulations. The relaxation time has been previously defined and the diffusion constant can be obtained from the mean squared displacement via the Einstein relation: $D = lim_{t\to\infty} < \Delta r^2(t)/6t >$. Both magnitudes present this power law dependence at moderate temperatures as it can be observed in Figure 6. The fits were performed fixing the glass temperature for both cases at $T^*_{MCT} = 0.105$. We also observe that the values of the exponents differ slightly: $\gamma_D = 1.9$ and $\gamma_\tau = 2.3$; discrepancy which has been previously found in other systems, for example in the Lennard-Jones mixture introduced by Kob et al.[34]. It is usually assumed that this discrepancy comes from the dynamical heterogeneities. Slow particles control the density correlation function, while the mean squared displacement is modulated by fast particles. Furthermore, it is worth noting that the power laws break down just at the lowest temperature investigated, that is, the closest to the glass transition. This fact is usually attributed to the presence of hopping processes which are present in the vicinity of the transition and restore ergodicity, the dynamical arrest is not complete and the divergence fails to describe these data. Additionally, the diffusion constant is represented in an Arrhenius plot in the inset to Fig. 6, where we observe that the colloidal electrolyte shows a non-Arrhenius behaviour.

As we mentioned above, MCT predicts a two-step relaxation process, where the decay from the plateau is given by the *von Schweidler* power-law series [32]:

$$F^s(q,t) = f^q_s - h^{s1}_q (t/\tau)^b + h^{s2}_q (t/\tau)^{2b} + O(t^{3b}). \quad (6)$$

Here $h^{s1}_q$ and $h^{s2}_q$ are amplitudes (which depend on the wave vector), $b$ is known as the *von Schweidler* exponent, and $f^s_q$ is the nonergodicity parameter. All of these parameters are not universal and depend on the driving mechanism for the glass transition. Under MCT, however, the exponent $b$ can be determined from $\gamma$, using the appropriate expressions [32]. We use $\gamma_\tau = 2.3$, yielding $b = 0.66$, because usually the value obtained via $\tau$-route is close to that one predicted by the theory [34]. In Figure 7, the correlation functions for different wavevectors at $\rho = 0.85$ and $T = 0.13$ are shown, where the fits to equation 6 are included as well. We can see that the fits describe correctly the decay of the correlation function from the plateau during one decade at high $q$ and two decades at low $q$. In addition, we have studied the wave vector dependence of the $\alpha$-relaxation time scale (inset in Fig. 7), where the time scale is defined as: $F^s_q(\tau^s_q) = f^s_q/e$. We can see that the relaxation is diffusive at low $q$ values, when $\tau_q \sim q^{-2}$, but crosses over to a different behaviour at high $q$, $\tau_q \sim q^{-1/b}$, as predicted by MCT.

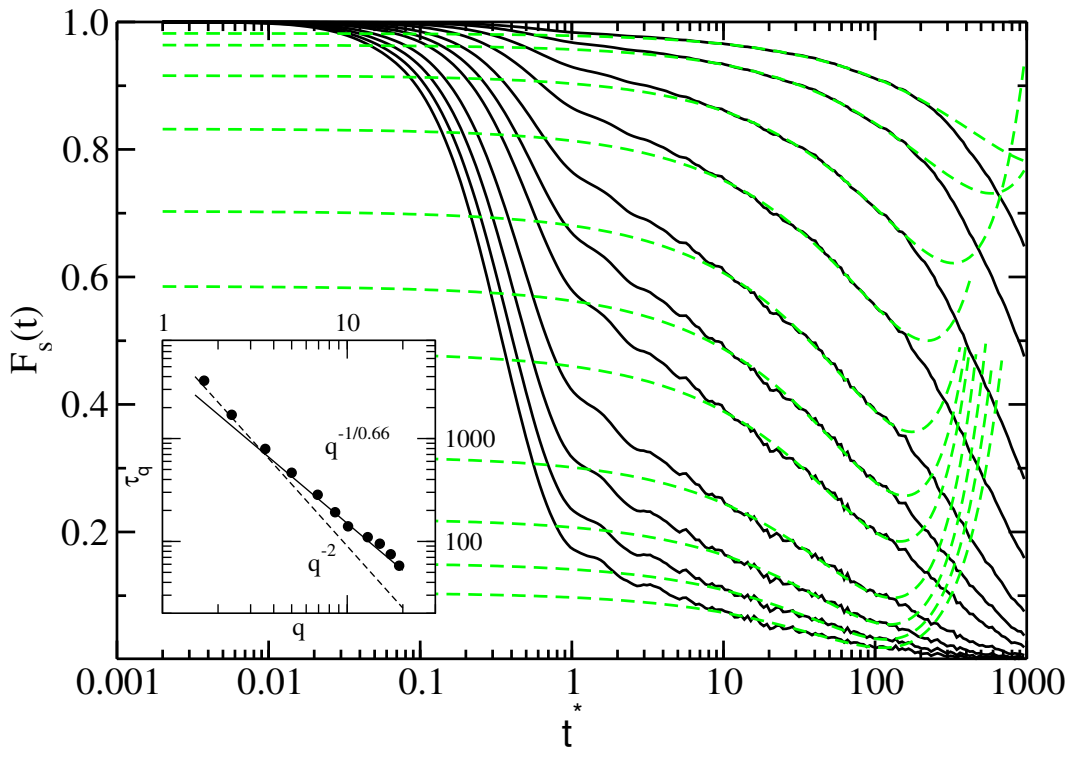

FIG. 7: *von Schweidler* fits at $\rho = 0.85$ and $T = 0.13$ with $\kappa\sigma = 6$ for different wave vector values (from top to bottom): $q\sigma$ = 1.68, 2.37, 3.6, 5., 6.92, 8.6, 10.1, 12.9, 15., 17.2, and 19.1. Inset: wave vector dependence of the $\alpha$-relaxation time scale, where $\tau_q$ is defined as $f^s_q(\tau_q) = f^s_q/e$.

To conclude the comparison with the MCT predictions, Figure 8 shows the wave vector dependence of the von Schweidler parameters $f^s_q$ and $h^{s1}_q$, obtained from the fittings in Fig. 7. The trends for these parameters are very similar to those reported for hard-spheres and Lennard-

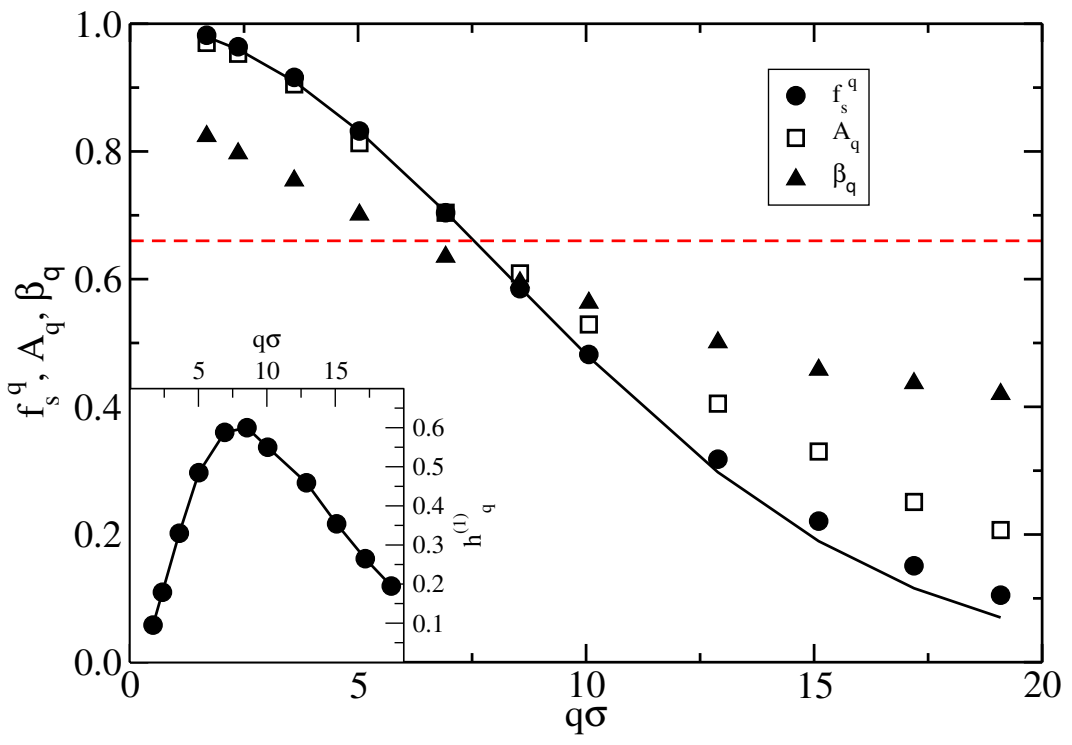

FIG. 8: Comparison between the *von Schweidler* and *Kohlrausch* parameters in the state point: $\rho = 0.85$ and $T = 0.13$ with $\kappa\sigma = 6$. The solid line is the Gaussian fit of the non ergodicity parameters. The dashed line marks the value of the *von Schweidler* exponent, $b$. Inset: wavevector dependence of the amplitude $h_q^{s1}$ in the *von Schweidler* fit.

| System | $\rho^*$ | $T^*_{glass}$ | $b$ | $r_l^2$ |
|---|---|---|---|---|
| $\kappa\sigma = 6$ | 0.90 | 0.139(5) | 0.66 | 0.037(2) |
| | 0.85 | 0.105(5) | 0.66 | 0.039(2) |
| $\kappa\sigma = 10$ | 0.90 | 0.148(7) | 0.65 | 0.035(2) |
| | 0.85 | 0.10(1) | 0.68 | 0.033(3) |
| $\kappa\sigma = 20$ | 0.90 | 0.155(3) | 0.66 | 0.033(2) |
| | 0.85 | 0.107(2) | 0.65 | 0.040(2) |
| Soft-Spheres | 1.134 | 4/9 | 0.53 | 0.0325 |

TABLE I: Summary of the most relevant MCT parameters for the colloidal electrolyte. Results for repulsive soft sphere ($u(r) = 1/r^{36}$) has been included for comparison (from ref. [48]).

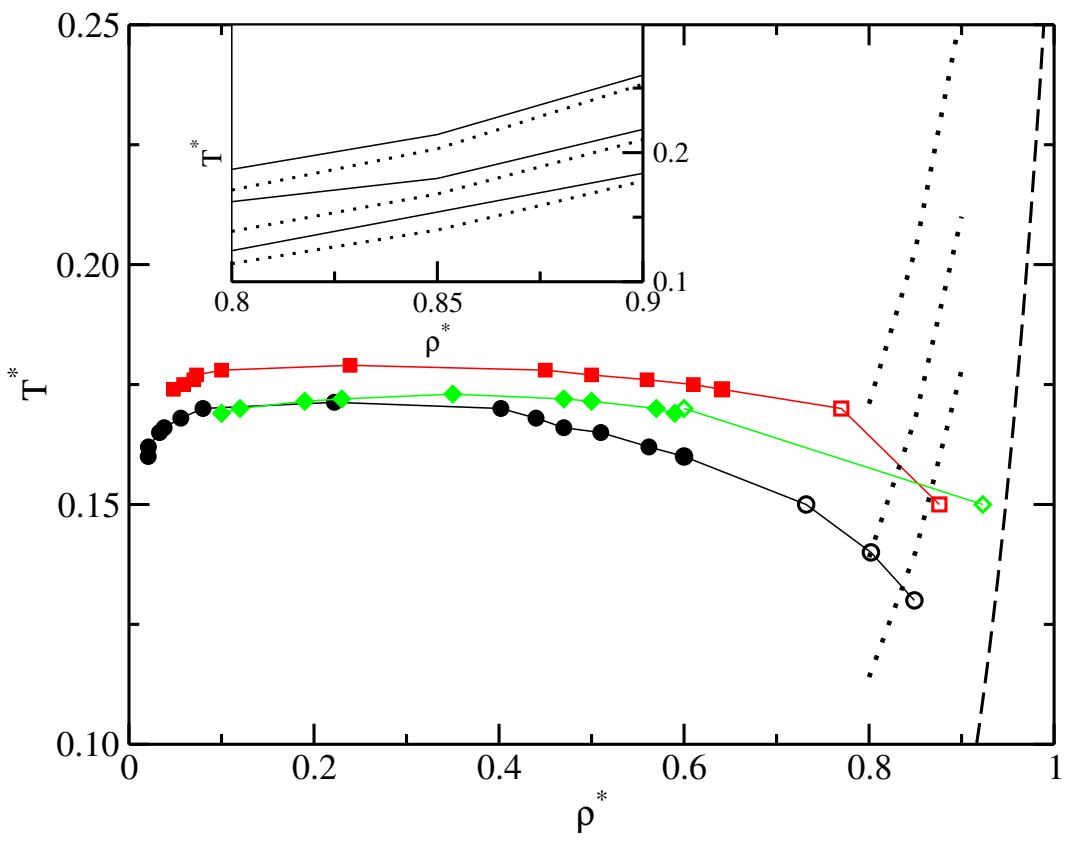

FIG. 9: Intercept of the glass line for soft sphere $u(r) = 1/r^{36}$ (dashed line) and the gas-liquid transition for the colloidal electrolytes: solid points from Ref. [9] and open symbols are results obtained using the method of the histograms (see text). Dotted lines are isodiffusivity lines for $\kappa\sigma = 6$: $D = 10^{-2.5}$, $D = 10^{-3.0}$, and $D = 10^{-3.5}$ (from top to bottom). Inset: isodiffusivity lines for $\kappa\sigma = 6$ (dotted lines) and $\kappa\sigma = 10$ (solid lines), from top to bottom (in both cases): $D = 10^{-2.5}$, $D = 10^{-3.0}$, and $D = 10^{-3.5}$. Results for $\kappa\sigma = 20$ are omitted for clarity.

Jones systems, where the $f_q^s$ is a decreasing function of the wave vector and the amplitudes show a peak at the wave vector where $S(q)$ presents its maximum (in our case at $q\sigma_{max} = 6.92$). At low $q$ the values of $f_q^s$ can be approximated by a Gaussian, $f_q^s \sim exp(-q^2 r_l^2/6)$ (solid line in this plot), $r_l$ being the localization distance. As we can see, such a fit produces good results, yielding $r_l^2 = 0.039$, value similar to that one found for the hard-sphere system, $r_l^2 = 0.034$. In this plot, parameters from the *Kolhrausch* fits are also included. $A_q^s$ shows similar values to $f_q^s$, as one expects by comparing equations 4 and 6. On the other hand, it is not obvious the relation between $b$ and $\beta_q$ because equation 4 is not an exact solution to MCT, but an expression which fits the MCT solutions. Nevertheless, it has been shown that in the limit $q \to \infty$ the expression 4 is an exact solution to the MCT equations [47] and then, $b$ should be equal to $\beta_{q\to\infty}$. Our results indicate that at low wave vector $\beta_q$ goes to one, that is, the decorrelation is diffusive, but, at long wave vector both exponent do not coincide ($\beta_{q\to\infty} < b$).

The results shown so far correspond to the isochore $\rho^* = 0.85$ for $\kappa\sigma = 6$. Similar findings were obtained for the other isochore as well as for $\kappa\sigma = 10$ and $\kappa\sigma = 20$. To summarize the comparison of the dynamics close to the glass transition with the universal and non universal predictions of the MCT for monocomponent systems, we provide the most relevant parameters in Table I, where the parameters for the repulsive part of the interaction potential are also provided, i.e. $u(r) = 1/r^{36}$ [48]. Note that the colloidal electrolyte tends to the repulsive soft potential $u(r) = 1/r^{36}$ at very high temperatures or when $\kappa\sigma \to \infty$. From this table we can see that in this region the glass transition for the colloidal electrolyte does not depend significantly on the interaction range: the *von Schweidler* exponents and the localization lengths are similar in all the cases. Also we have found similitudes with the repulsive soft potential, though $b$ presents higher values for the colloidal electrolyte what means that the decay from the plateau is faster in ionic colloids. Aside the similitudes, we should be aware that both liquids (the repulsive soft-spheres and the colloidal electrolyte liquid) present an important difference, namely, in the soft repulsive potential each liquid particle is surrounded approximately by 12 nearest neighbours, while 8 oppositely charged particles cage each particle in the colloidal electrolytes (recall that it forms a CsCl crystal (BCC) [10, 11]).

This MCT analysis points out the possibility that the glass line is rather similar for all the interaction ranges. Under MCT, the glass transition can be considered as the locus in the $T^* - \rho^*$ plane of zero diffusion constant, that is, the zero isodiffusivity line. Thereby, in the inset to Fig. 9 the isodiffusivity lines are shown for different interaction ranges, showing that they do not change sig-nificantly with the interaction range and then, supporting the idea that the glass line in this region is almost independent of the attractive interactions. Additionally, the properties of the glass transition in colloidal electrolytes is similar to those of inverse power potential with $n = 36$ (see table I), i.e. the repulsive part of the total interaction in eq. 1. Therefore, we will approximate the glass line of the colloidal electrolyte by the glass transition of the repulsive system $u(r) = 1/r^{36}$ [48]. For inverse power potentials, density and temperature are coupled by means of a dimensionless parameter $\Gamma = \rho^*/T^{*1/n}$ and therefore, the properties computed along an isochore or isotherm are sufficient in order to determine the complete phase behaviour of these systems [49]. In Figure 9 we represent the gas-liquid coexistence points for the colloidal electrolytes and the glass line for the inverse power potential with $n = 36$ [48]. First, observe that the glass line matches the extrapolation of the isodiffusivity lines for $\kappa\sigma = 6$ (included in this plot). Second, the results obtained in this way are consistent with those of Fig. 3 as far as the crossing points between the binodal and the glass line are reproduced. Consequently, we can conclude that in this region the colloidal electrolyte presents a repulsive glass which is driven mainly by the steric interactions, the effect of the attractions being negligible. This repulsive glass arrests the gas-liquid transition at moderate temperatures.

Interestingly, we have not found any signature of reentrant behaviours in the colloidal electrolyte as measured by computing the isodiffusivity lines for $\kappa\sigma = 6$, $\kappa\sigma = 10$, and $\kappa\sigma = 20$ in the ergodic region (see Fig. 9 and its inset). In monocomponent systems, the reentrant glass phenomenon is associated with short ranged attractions (the particles that cage a central particle attract each other forming holes by which the central particle can escape), which also are responsible for attractive glasses and gelation at lower density[16, 18, 28]. In fact, this re-entrance region connects the attractive and repulsive glasses. Here, due to correlations between oppositely charged colloids each particle is caged by oppositely charged colloids and the interaction between particles that cage a central particle would be repulsive. If so, the re-entrance could disappear. Although we have not found any signatures of attraction driven glasses, it is intriguing that at low temperatures gels were indeed present and had a remarkable effect in the liquid-vapour phase separation. How these gels are connected with the glass line here studied, driven by repulsions, and the presence of attractive glasses deserve further study. In particular, it would be valuable MCT results for the colloidal electrolyte to compare with the results of this work.

## IV. CONCLUSIONS

In this paper, the low temperature behaviour and the dynamics close to the glass line has been studied by means of Molecular Dynamics simulations for the symmetrical colloidal electrolyte, where the electrostatic interactions were modeled via the DLVO effective interactions. Simulations along two isochores show a rich behaviour. At moderate temperatures, below the critical temperature, vapour-liquid coexistence is observed. At lower temperatures, two different mechanisms have been identified to arrest the gas-liquid separation: vitrification of the liquid due to the crossing of the liquid branch with a glass line, and gelation at very low temperatures, identified by a ramified structure that contains all of the particles in the system. This scenario is independent of the interaction range and can explain recent experimental results in a similar system [7].

Moreover, the dynamics of this model system near the crossing point between the binodal and the glass line has been studied and compared with the universal predictions of the ideal MCT for monocomponent systems. Indeed, the theoretical predictions describe correctly many of the simulation results in a wide temperature range in the same way as for other glass-forming liquids, although they fail just at the transition. Interestingly, the dynamical properties of this complex system do not depend significantly on the interaction range and also they are similar to those of inverse power potential with $n = 36$, i.e. the repulsive part of the total interaction (see eq. 1). This fact indicates that in this region the glass transition is driven mainly by the steric interactions. Finally, the isodiffusivity lines do not show the re-entrant glass phenomena, usual in monocomponent systems with short range attractions, and where gelation is also present.

## V. ACKNOWLEDGMENTS

We warmly thank M.E. Cates for useful and stimulating discussions. J.B.C. acknowledges the hospitality during his stay in the School of Physics (University of Edinburgh), when part of these results were obtained. This work is financially supported by the *Ministerio de Educación y Ciencia*, under project number MAT2004-03581.